\documentclass[12pt]{article}
\usepackage{graphicx}
\usepackage{color}

\usepackage{amsmath} %
\usepackage{amsfonts}

\newcommand{\sech}{{\rm \,sech}}

\begin{document}

\title{Dynamics of internal envelope solitons in a rotating fluid of a variable depth}

\author{Y.~A.~Stepanyants \\
{\small Faculty of Health, Engineering and Sciences, University of Southern Queensland,}\\
{\small Toowoomba, QLD, 4350, Australia, e-mail: {\color{blue} Yury.Stepanyants@usq.edu.au}} \\
{\small and} \\
{\small Department of Applied Mathematics, Nizhny Novgorod State Technical, University,}\\
{\small Nizhny Novgorod, 603950, Russia.}}

\maketitle

{\centerline {\bf Abstract}}

We consider the dynamics of internal envelope solitons in a two-layer rotating fluid with a linearly varying bottom. It is shown that the most probable frequency of a carrier wave which constitutes the solitary wave is the frequency where the growth rate of modulation instability is maximal. An envelope solitary wave of this frequency can be described by the conventional nonlinear Schr\"odinger equation. A soliton solution to this equation is presented for the time-like version of the nonlinear Schr\"odinger equation. When such envelope soliton enters a coastal zone where the bottom gradually linearly increases, then it experiences an adiabatical transformation. This leads to an increase of soliton amplitude, velocity, and period of a carrier wave, whereas its duration decreases. It is shown that the soliton becomes taller and narrower. At some distance it looks like a breather, a narrow nonstationary solitary wave. The dependences of soliton parameters on the distance when it moves towards the shoaling are found from the conservation laws and analysed graphically. Estimates for the real ocean are presented.\\


\section{Introduction}
\label{Sect01}

The effect of Earth' rotation on the dynamics of nonlinear waves in the oceans has been extensively studied in previous decades (see, for example, Refs. \cite{Ostrovsky78, Grimshaw85, GrimOstrShrStep, Grimshaw02, Helfrich06, Grimshaw14, Ostrovsky15, Step19} and references therein). As is well-know, wave propagation in big lakes can be also affected by the Earth's rotation \cite{Boegman10, Fuente10, Preusse12, Ulloa15, Rojas18}. In particular, the dynamics of solitary waves has been investigated within the framework of the Ostrovsky equation and it has been established that they cannot propagate steadily due to the permanent radiation of small-amplitude long waves \cite{Leonov, Galk&Step} (they can steadily propagate, however, being supported by a long background wave \cite{Gilman95, OstrovStep16}). As a result, an initial solitary wave experiences a terminal decay which leads formally to it vanishing in a finite time \cite{GrimHeOstr, GrimOstrShrStep}. However, the process of solitary wave decay is more complicated in reality and leads eventually to the formation of envelope solitons described by the nonlinear Shr\"odinger (NLS) equation or its modifications \cite{Grimshaw08, Grimshaw12, Grimshaw13, Grimshaw16, wj14, wj15}. In an inhomogeneous medium the dynamics of solitary waves is determined by synergetic effects of inhomogeneity and fluid rotation. In particular, at a certain relationship between these two factors, a Korteweg--de Vries (KdV) soliton propagating towards a coast with a gradually decreasing depth can preserve its shape and amplitude, whereas its width and velocity adiabatically change \cite{Step19}.

In the real ocean when a KdV soliton approaches a coastal zone it can experience a terminal decay in the domain where the depth is constant, so that it can be ultimately transformed into an NLS envelop soliton, and then the envelop soliton can enter into the inhomogeneous domain where an oceanic depth gradually decreases. An NLS soliton can be formed from a rather arbitrary initial perturbation independently on the transformation of a KdV soliton. It is a matter of interest to study the adiabatic dynamics of an NLS envelop soliton when it approaches a shoaling zone. To this end we consider below different models of NLS-type equations for water waves in a rotating ocean, their solutions in the form of envelop solitons, and the dynamics of such solitons in the ocean with a gradually decreasing depth.

\section{The variable coefficients Ostrovsky equation}
\label{Sect02}

The dynamics of weakly nonlinear waves in a rotating inhomogeneous ocean can be described by the Ostrovsky equation with the variable coefficients (see, e.g., Ref. \cite{Grimshaw02, Grimshaw14} and references therein):
\begin{equation}%
\label{vcrKdV}%
\frac{\partial}{\partial x}\left(\frac{\partial u}{\partial t} + c\frac{\partial u}{\partial x} + \alpha u\frac{\partial u}{\partial x} + \beta \frac{\partial^3 u}{\partial x^3} +
\frac{u}{2}\frac{dc}{dx}\right) = \gamma u\,,
\end{equation}
where for internal waves in a two-layer fluid in the Boussinesq
approximation the coefficients are:
\begin{equation}%
\label{Intern}%
c = \sqrt{g\frac{\delta \rho}{\rho}\frac{h_1h_2}{h_1 + h_2}} \quad
\alpha = \frac{3}{2}c\frac{h_1 - h_2}{h_1h_2}, \quad \beta =
\frac{c}{6}h_1h_2, \quad \gamma = \frac{f^2}{2c},
\end{equation}
where $h_1 = const$, and $h_2 = h_2(x)$.

For the boundary-value problem, $u(t, 0) = f(t)$, this equation
can be presented in the alternative (``signalling'') form dubbed here the time-like
Ostrovsky equation:
\begin{equation}%
\label{bvrKdV}%
\frac{\partial}{\partial t}\left(\frac{\partial u}{\partial x} + \frac{1}{c(x)}\frac{\partial u}{\partial t} - \frac{\alpha}{c^2(x)} u\frac{\partial u}{\partial t} - \frac{\beta}{c^4(x)} \frac{\partial^3 u}{\partial t^3} +
\frac{u}{2c(x)}\frac{dc}{dx}\right) = -\gamma(x) u.
\end{equation}

The dispersion relation corresponding to Eq. (\ref{bvrKdV}) with the constant coefficients is:
\begin{equation}%
\label{bvDisp}%
k(\omega) = \frac{\omega}{c} + \frac{\beta}{c^4}\omega^3 -
\frac{\gamma}{\omega}.
\end{equation}

In the homogeneous nonrotating ocean Eq.~(\ref{bvrKdV}) has a particular solution in the form of a KdV soliton:
\begin{equation}%
\label{KdV-Sol}%
u(t, 0) = A_{K}\sech^2\left(\frac{t - x/V}{\Delta_t}\right),
\end{equation}
where the soliton temporal duration $\Delta_t$ and speed $V_{K}$ are related to the soliton amplitude $A_{K}$ through the formulae:
\begin{equation}%
\label{EnBalEq}%
\Delta_t
= \frac{1}{c}\sqrt{\frac{12\beta}{\alpha A_K}}, \quad V_K
= \frac{c}{1 - \alpha A_K/3c}.
\end{equation}

It is assumed that such a soliton in the course of propagation in a rotating ocean of a constant depth after long-term evolution was ultimately transformed into an envelope soliton as described in Refs. \cite{Grimshaw08, Grimshaw12, Grimshaw13, Grimshaw16}. We assume then that the resultant envelope soliton enters into the coastal zone, where the depth gradually decreases with $x$. Our aim is to describe the fate of such a soliton and present the dependences of its parameters on distance.

\section{The variable coefficients NLS equation}
\label{Sect03}

In the coastal zone where the bottom profile gradually varies with the distance, the NLS equation describing soliton evolution contains the additional ``inhomogeneous'' term and has the form of the ``time-like NLS'' (TNLS) equation (cf. \cite{Djordjevic78, Benilov05, Benilov06}):
\begin{equation}%
\label{VarCoefNLS}%
i\left(\frac{\partial \psi}{\partial x} + \frac{1}{c_g(x)}\frac{\partial \psi}{\partial t} +
\frac{\psi}{2}\frac{d\ln{c_g}}{dx}\right) + \frac{p(x)}{c^3_g(x)}\frac{\partial^2 \psi}{\partial t^2}
 + \frac{q(x)}{c_g(x)} |\psi|^2 \psi = 0,
\end{equation}
where the coefficients $p(x)$ and $q(x)$ are linked with the
coefficients of the Ostrovsky equation (\ref{vcrKdV}) (cf.
\cite{Grimshaw16}):
\begin{equation}%
\label{CoefNLSeq}%
p(x) = -c^3(x)\left[\frac{3\beta(x)}{c^4(x)}\omega -
\frac{\gamma(x)}{\omega^3}\right], \quad q(x) = -\frac 23
\frac{\alpha^2(x)\omega^3c(x)} {4\beta(x)\omega^4 + \gamma(x)c^4(x)}.
\end{equation}

The group velocity $c_g(x)$ as follows from the Ostrovsky equation
(\ref{vcrKdV}) is
\begin{equation}%
\label{GroupVel}%
c_g(x) = c(x) - 3\beta(x) k^2(x) - \frac{\gamma(x)}{k^2(x)}.
\end{equation}
Note that in a stationary, but spatially inhomogeneous media, the
wavenumber $k$ depends on $x$, whereas the frequency conserves.

The dependence of the wavenumber of a carrier wave on the spatial
coordinate $x$ follows from the dispersion relation
(\ref{bvDisp}) where $\omega = const.$, and other parameters, $c$, $\beta$, and $\gamma$ depend on $x$.

Using the results obtained in \cite{Grimshaw16}, let us present
the soliton solution of TNLS equation (\ref{VarCoefNLS})
assuming that all its coefficients are constants:
\begin{equation}%
\label{EnvSol}%
\psi = A\sech{\left[\Omega\left(t - \frac{x}{c_g} -
\frac{x}{V}\right)\right]}e^{i\left[\sigma\left(t - x/c_g\right) -
\kappa x\right]},
\end{equation}
where the amplitude $A$ and speed $V$ can be considered as the free parameters, whereas soliton duration $\Delta T = 1/\Omega$, $\Omega = Ac_{g0}\sqrt{q_0/2p_0}$, and the chirp and gauge respectively are:
\begin{equation}%
\label{EnvSolPar}%
\sigma = \frac{c^3_{g0}}{2p_0V}, \quad \kappa = \frac{q_0}{2c_{g0}}\left(\frac{c^4_{g0}}{2p_0q_0V^2} - A^2\right).
\end{equation}

In the particular case when $V \to \infty$, we obtain
\begin{equation}%
\label{EnvSolPar0}%
\Omega = Ac_{g0}\sqrt{\frac{q_0}{2p_0}}, \quad \sigma = 0, \quad
\kappa = -\frac{q_0A^2}{2c_{g0}}.
\end{equation}

Solution (\ref{EnvSol}) is invalid when the dispersion coefficient
$p$ in the TNLS equation (\ref{VarCoefNLS}) vanishes. This occurs
at the frequency $\omega_{m0} = c_0\sqrt[4]{\gamma_0/3\beta_0}$
when the group velocity has a maximum $\left(c_{g0}\right)_{max} = c_0
-2\sqrt{3\beta_0\gamma_0}$. In a such case the generalised NLS
equation derived in Refs. \cite{Grimshaw08, Grimshaw12} should be
used.

If the envelope soliton (\ref{EnvSol}) enters the region where the
coefficients of TNLS equation (\ref{VarCoefNLS}) gradually varies
with $x$, then the adiabatic evolution of the soliton main parameters
$A$ and $V$ can be determined from the balance equations, following from the conservation laws for the TNLS equation
(\cite{Andronov, Churilov}, see also \cite{Stepanyants92}).
Alternatively, the rigorous asymptotic theory can be developed,
but as has been shown in Ref. \cite{Karpman79}, the outcome
reduces to the first two conservation laws, the conservation of
the total flux of ``quasi-particles'':
\begin{equation}%
\label{PartNumb}%
N \equiv c_g(x)\int\limits_{-\infty}^{+\infty}|\psi|^2dt = const
\end{equation}
and the conservation of quasi-momentum:
\begin{equation}%
\label{QuasiMom}%
P \equiv \frac{2i}{c_g(x)} \int\limits_{-\infty}^{+\infty}
\psi^*\psi_t\,dt = const.
\end{equation}

Substituting the soliton solution (\ref{EnvSol}) into Eq.
(\ref{PartNumb}), we obtain after simple manipulations:
\begin{equation}%
\label{Eq1}%
\frac{A}{A_0} = \sqrt{\frac{p(x_0)}{p(x)}\frac{q(x)}{q(x_0)}}.
\end{equation}

Then from Eq. (\ref{QuasiMom}) using Eq. (\ref{EnvSol}) and the result
obtained for soliton amplitude in Eq. (\ref{Eq1}), we derive for the soliton speed:
\begin{equation}%
\label{Eq2}%
\frac{A(x)^2\sigma}{c_g(x)} \sim \frac{A(x)^2c^2_g(x)}{V(x)p(x)} =
const.
\end{equation}

From here we find:
\begin{equation}%
\label{Eq3}%
\frac{V(x)}{V(x_0)} = \frac{c^2_g(x)}{c^2_g(x_0)}
\frac{A^2(x)}{A^2(x_0)} \frac{p(x_0)}{p(x)} =
\frac{c^2_g(x)}{c^2_g(x_0)} \frac{p^2(x_0)}{p^2(x)}
\frac{q(x)}{q(x_0)}.
\end{equation}

After that we can determine the evolution of other soliton parameters
by means of relationships (\ref{EnvSolPar}).

In the particular case when $V \to \infty$ $(\sigma \to 0)$, Eq.
(\ref{Eq3}) vanishes, and we have only one equation (\ref{Eq1})
determining the evolution of soliton amplitude.

As mentioned above, the TNLS equation (\ref{VarCoefNLS}) becomes
invalid when its dispersion coefficient vanishes, then the
generalised NLS equation derived in Refs. \cite{Grimshaw08,
Grimshaw12} should be used. In the next section we consider this
special case.

\section{Generalised variable coefficients NLS equation}
\label{Sect04}

In the vicinity of the point where the second-order dispersion in
the TNLS equation becomes very small or vanishes, i.e., when $p(x)
\to 0$, the equation should be generalised by inclusion of additional
terms \cite{Grimshaw08}. For the boundary-value problem the
corresponding equation reads:
$$
i\left(\psi_x + \frac{1}{c_g(x)}\psi_t +
\frac{\psi}{2}\frac{d\ln{c_g}}{dx}\right) + \frac{p(x)}{c^3_g(x)}
\psi_{tt} + \frac{q(x)}{c_g(x)} |\psi|^2 \psi - {}
$$
\begin{equation}%
\label{GenNLS}%
i\frac{\nu(x)}{c_g^4(x)}\psi_{ttt} -
\frac{i}{c_g^2}r_1(x)\left[\psi^2\bar\psi_t + r_2(x)|\psi|^2\psi_t\right] = 0,
\end{equation}
where $\bar\psi$ stands for complex-conjugate with respect to
function $\psi$ and the coefficients $\nu(x)$ and $r(x)$ are linked with the
coefficients of the Ostrovsky equation (\ref{vcrKdV}) (cf. \cite{Grimshaw08}): 
$$
\nu(x) = -\left[\beta(x) + \frac{\gamma(x) c^4(x)}{\omega^4}\right],
$$
$$
r_1(x) = \frac 23\frac{\alpha^2(x)\omega^2c^2(x)}{
	4\beta(x)\omega^4 + \gamma(x)c^4}, \quad r_2(x) = \frac{4\beta(x)\omega^4 + 5\gamma(x)c^4(x)}{4\beta(x)\omega^4 + \gamma(x)c^4(x)}.
$$

The soliton solution to Eq. (\ref{GenNLS}) with the constant
coefficients has the same form as Eq. (\ref{EnvSol}), but in
contrast to the conventional NLS soliton, it is now a one
parametric solution with only one independent parameter. If we
chose the amplitude as the independent parameter, then other
soliton parameters can be presented as follows:
\begin{equation}%
\label{EnvOmega}%
\Omega = Ac_g\sqrt{\frac{r_1(1+r_2)}{6\nu}}, \quad \sigma = \frac{c_gq}{2r_1}\left[1 - \frac{pr_1(1+r_2)}{3q\nu}\right], 
\end{equation}
\begin{equation}%
\label{EnvSpeed}%
V = \frac{12c_g^2\nu r_1^2}{9q^2\nu^2 + 6pq\nu r_1(1-r_2) - 3p^2r_1^2 - p^2r_1^2r_2(2 - r_2) - 2A^2\nu r_1^3(1 + r_2)}, 
\end{equation}
\begin{equation}%
\label{EnvKappa}%
\kappa = \frac{-18A^2\nu(1 + r_2)\left[pr_1(1 - r_2) + 3q\nu\right] + p^3\left[5 + 9r_2 + 3r_2^2 - r_2^3\right] + F}{216c_g\nu^2},
\end{equation}
$$
\mbox{where} \quad F = \frac{9pq\nu}{r_1^2}\left[3q\nu\left(1 - r_2\right) - pr_1\left(3 + 2r_2 - r_2^2\right)\right] + 27\left(\frac{q\nu}{r_1}\right)^3.
$$

This solution can be reduced to the conventional NLS envelope
soliton (\ref{EnvSol}) if we assume that $\nu \to 0$, $r_1 \to 0$, $r_2 \to 0$,
but such that $r_1(1+r_2)/\nu \to 3q/p$.

In another limiting case when $p = 0$, the parameters of the
soliton solution as the functions of amplitude are:
\begin{equation}%
\label{EnvAmp2}%
V = \frac{12c_g^2\nu r_1^2}{\nu\left[9q^2\nu - 2A^2 r_1^3\left(1 + r_2\right)\right]}, \quad \sigma = \frac{c_gq}{2r_1}, \quad \kappa =
\frac{q\left[\nu q^2 - 2A^2r_1^3\left(1 + r_2\right)\right]}{8c_gr_1^3},
\end{equation}
and $\Omega(A)$ remains the same as in Eq. (\ref{EnvOmega}).

When a solition propagates in the inhomogeneous medium, its
frequency remains constant, but the wavenumber varies in
accordance with the formula (cf. Eq. (\ref{bvDisp})):
\begin{equation}%
\label{VarWaveNumb}%
k(x) = \frac{\omega}{c(x)} + \frac{\beta(x)}{c^4(x)}\omega -
\frac{\gamma(x)}{\omega}.
\end{equation}

As follows from this equation, the critical wavenumber, where the
group velocity has a maximum, adiabatically changes in accordance
with the variation of parameters. Therefore, if the envelope
soliton has been created near the critical point, it will remain
further at the corresponding critical point.

According to the numerical results of Ref. \cite{Grimshaw08},
envelope solitons emerging from the KdV solitons in the course of
long-term evolution have almost zero correction to the group speed
(see Fig. 9 in \cite{Grimshaw08}). In our notations this
corresponds to $V = \infty$. Such value of speed correction corresponds to a soliton with a fixed amplitude:
\begin{eqnarray}
A_0 &=& \sqrt{\frac{3\nu q\left[3\nu q + 2pr_1(1-r_2)\right] - p^2r_1^2\left[3 + r_2(2 - r_2)\right]}{2\nu r_1^3(1 + r_2)}} \quad \mbox{for Eq. (\ref{EnvSpeed})}, \label{A01} \\
A_0 &=& \sqrt{\frac{9\nu q^2}{2r_1^3\left(1 + r_2\right)}}\quad \mbox{for Eq. (\ref{EnvAmp2})}. \label{A02}
\end{eqnarray}

The gauge $\kappa$ and parameter $\Omega$ determining the half-width of a soliton are also fixed in this case:
\begin{eqnarray}
\kappa_0 &=&
\frac{\left[pr_1\left(1 + r_2\right) - 3q\nu\right]\left[pr_1\left(2 - r_2\right) + 3q\nu\right]^2}{27c_g\nu^2 r_1^3}
\quad \mbox{for Eq. (\ref{EnvSpeed})}, \label{K01} \\
\kappa_0 &=& -\frac{\nu q^3}{c_gr_1^3} \quad \mbox{for Eq.
(\ref{EnvAmp2}) \label{K02}};
\end{eqnarray}
\begin{eqnarray}
\Omega_0 &=& \frac{2c_g}{r_1\nu}\sqrt{3\nu q\left[\nu q + 2pr_1(1-r_2)\right] - p^2r_1^2\left[1 + r_2(2 - r_2)/3\right]}, \quad \mbox{for Eq. (\ref{EnvSpeed})}, \label{Om1} \\
\Omega_0 &=& \frac{qc_g\sqrt{3}}{2r_1} \quad
\mbox{for Eq. (\ref{EnvAmp2}) \label{Om2}}.
\end{eqnarray}

In an inhomogeneous medium all soliton parameters (\ref{EnvOmega})--(\ref{EnvKappa}) vary with $x$. The
equation governing the parameter variations follows from the conservation of a total flux of ``quasi-particles'' as per Eq. (\ref{PartNumb}):
\begin{equation}%
\label{AmplVar}%
\frac{A(x)}{A(x_0)} = \sqrt{\frac{\nu(0)}{\nu(x)}\frac{r_1(x)\left[1 + r_2(x)\right]}{r_1(0)\left[1 + r_2(0)\right]}}.
\end{equation}

Then using Eqs. (\ref{EnvOmega})--(\ref{EnvKappa}), one can
determine variation of the parameter $\Omega(x)$, velocity $V(x)$, and gauge $\kappa(x)$, whereas the chirp $\sigma(x)$ varies adiabatically as per Eq. (\ref{EnvOmega}) and independently of soliton amplitude. 

In the limiting case when $p = 0$, all soliton parameters are determined only by the coefficients of generalised NLS equation (\ref{GenNLS}). Then its amplitude can vary adiabatically with $x$ only at a very special relationship between the coefficient $\alpha$ and linear wave speed $c$ such that $\alpha^2(x) c(x) = const$.

\section{What is the most probable frequency of envelope soliton?}
\label{Sect05}

Consider the constant coefficients NLS equation (\ref{VarCoefNLS}). As follows from the analysis of stability of a uniform wave train with the amplitude  $a_0$, the modulation instability occurs when $pq > 0$. In our case, as follows from Eq.~(\ref{CoefNLSeq}), $q$ is always positive, whereas $p$ becomes positive when $\omega > \omega_{m0}$. The maximum of growth rate of modulation instability is $\Gamma = qa_0^2/c_g$ (see, e.g. \cite{Karpman73, OstrPotap99, ZakhOstr09}). Substituting here $q$ as per Eq.~(\ref{CoefNLSeq}), we obtain:
\begin{equation}%
\label{Gamma}%
\Gamma(\omega) = \frac 23\frac{(\alpha a_0)^2\omega^3}{4\beta\omega^4 + 3\gamma c^4}.
\end{equation}
This expression has a maximum at $\omega_{max} = c(9\gamma/4\beta)^{1/4}$, where the maximal growth rate for the given amplitude of a wave train is:
\begin{equation}%
\label{GammaMax}%
\Gamma_{max} = \frac{(\alpha a_0)^2}{4\sqrt{6}\beta c}\sqrt[4]{\frac{\beta}{\gamma}}.
\end{equation}

Thus, one can expect that an envelope soliton will evolve from a quasilinear wave train with the carrier frequency $\omega_{max}$ corresponding to the maximal growth rate of modulation instability. This agrees with the arguments presented in Ref. \cite{wj15} where it was shown that an envelope soliton cannot emerge at the carrier frequency $\omega_{m0}$, as was assumed in the papers \cite{Grimshaw08, Grimshaw12, Grimshaw13}, but should emerge at a higher carrier frequency. The concrete carrier frequency was not found, but only roughly estimated from the numerical data. As follows from the above theory, the relative frequency shift is fairly significant:
\begin{equation}%
\label{RelFreqShift}%
\frac{\omega_{max} - \omega_{m0}}{\omega_{m0}} = \sqrt[4]{\frac{27}{4}} - 1 \approx 0.62.
\end{equation}

Therefore, one can expect that in the process of evolution of a KdV soliton in a uniform rotating ocean it eventually transforms into an NLS envelope soliton with the carrier frequency $\omega_{max}$. If a such soliton enters into a coastal zone with a gradually decreasing depth, it then changes adiabatically, and its basic parameters, amplitude $A$ and velocity $V$ vary with $x$ in accordance with Eq. (\ref{Eq1}) and (\ref{Eq3}).

\section{Estimations for the real oceanic conditions}
\label{Sect06}

\begin{figure}[h!]
	{\centerline{\includegraphics [scale=0.6]{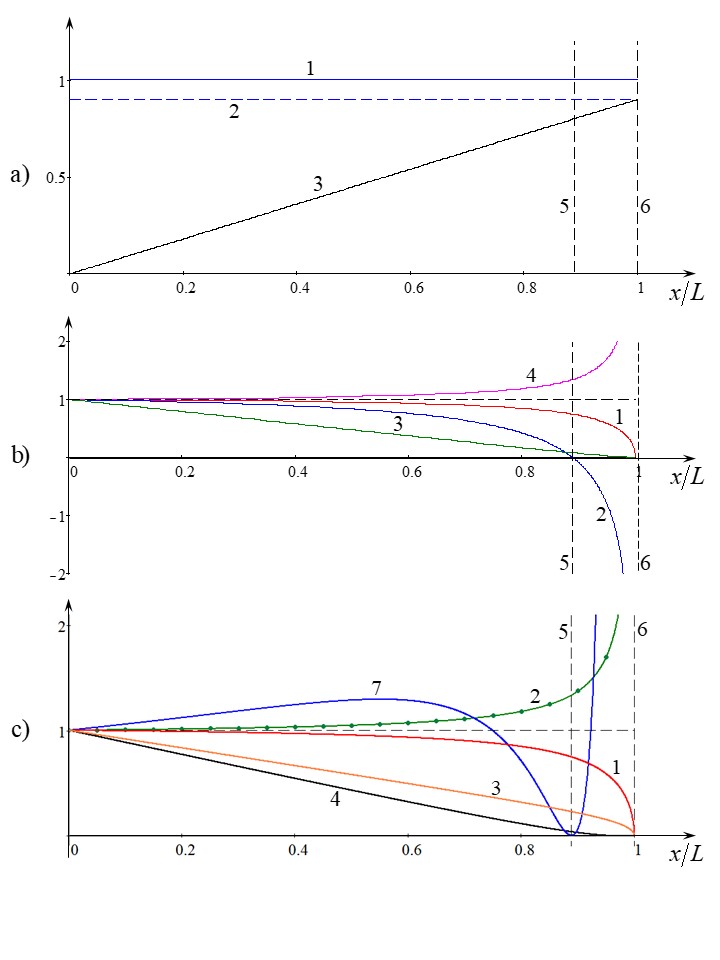}}} %
\vspace{-1.5cm}%
	\caption{\small Frame a): line 1 -- free surface in the basin of the total depth $H(0) = 500$ m; line 2 -- position of a pycnocline at the depth $h_1 = 50$ m; line 3 -- normalised bottom profile as a function of $x/L$. Frame b): Normalised coefficients of the time-like Ostrovsky equation (\ref{bvrKdV}): line 1 -- $c(x)/c(0)$; line 2 -- $\alpha(x)/\alpha(0)$; line 3 -- $\beta(x)/\beta(0)$; line 4 -- $\gamma(x)/\gamma(0)$; line 5 -- vertical line showing the distance where the upper and lower layers become equal; line 6 -- vertical line showing the distance where the thickness of the lower layer vanishes. Frame c): Normalised coefficients of TNLS equation (\ref{VarCoefNLS}): line 1 -- $c(x)/c(0)$; line 2 -- $k(x)/k(0)$; line 3 -- $c_g(x)/c_g(0)$; line 4 -- $p(x)/p(0)$; line 7 -- $q(x)/q(0)$.} %
	\label{f01}
\end{figure}

Let us assume that in the coastal zone the bottom profile is a linear function of a distance, so that the depth linearly decreases from $H_1 = 500$ m up to $H_2 = 50$ m at the distance $L = 10^6$ m, and the pycnocline is located at the depth $h_1 = 50$ m as shown in Fig. \ref{f01}, frame a). Then the initial thickness of the lower layer $h_2(0) = 450$ m; it gradually decreases with the distance $h_2(x) = h_2(0)(1 - x/L)$ and turns to zero at $x = L$. The normalised bottom profile $1 - \left[h_1 + h_2(x)\right]/H_1$ is shown by line 3 in frame a) of Fig. \ref{f01}. Let us set the Coriolis parameter $f = 10^{-4}$ 1/s which is a typical value for the moderate latitudes, and $g' \equiv g\Delta \rho/\rho = 0.03$ m/s$^2$. With these parameters we obtain for the time-like Ostrovsky equation (\ref{bvrKdV}) the following values of coefficients: \\
$c = 1.162$ m/c, $\alpha = -3.1\cdot 10^{-2}$ 1/s, $\beta = 4.36\cdot 10^{3}$ m$^3$/s, $\gamma = 4.3\cdot 10^{-9}$ 1/(m$\cdot$ s).\\
The dependences of these coefficients on the distance are shown in Fig. \ref{f01}, frame b). The nonlinear coefficient $\alpha$ vanishes at some distance $x_1$, where $h_2(x_1)$ becomes \linebreak
\clearpage
equal to $h_1$. Then it changes its sign and becomes positive. This effect is well-known (see, e.g., \cite{Apel07} and references therein). (To avoid confusion, it should be borne in mind that the nonlinear coefficient in this figure is presented in the normalised form $\alpha(x)/\alpha(0)$, therefore this ratio is positive when $x = 0$ and becomes negative when $x_1 < x < L$.)

Frame c) demonstrates the dependences of coefficients in the TNLS equation (\ref{VarCoefNLS}). Line 1 for the normalised linear speed $c(x)$ in this frame is the same as in frame b), and line 3 illustrates the dependence of normalised group speed $c_g(x)$ as per Eq.~(\ref{GroupVel}). Line 2 shows the dependence of the carrier wave number for the envelope soliton (\ref{EnvSol}) when $\omega = \omega_{max}$. This dependence is practically indistinguishable from the similar dependence plotted for $\omega = \omega_{m0}$; the latter is shown in the same frame by dots on line 2. Line 4 shows the dependence of dispersion coefficient $p(x)$, and line 7 -- the dependence of nonlinear coefficient $p(x)$ as per Eq. (\ref{CoefNLSeq}). 

\begin{figure}[h!]
	{\centerline{\includegraphics [scale=0.7]{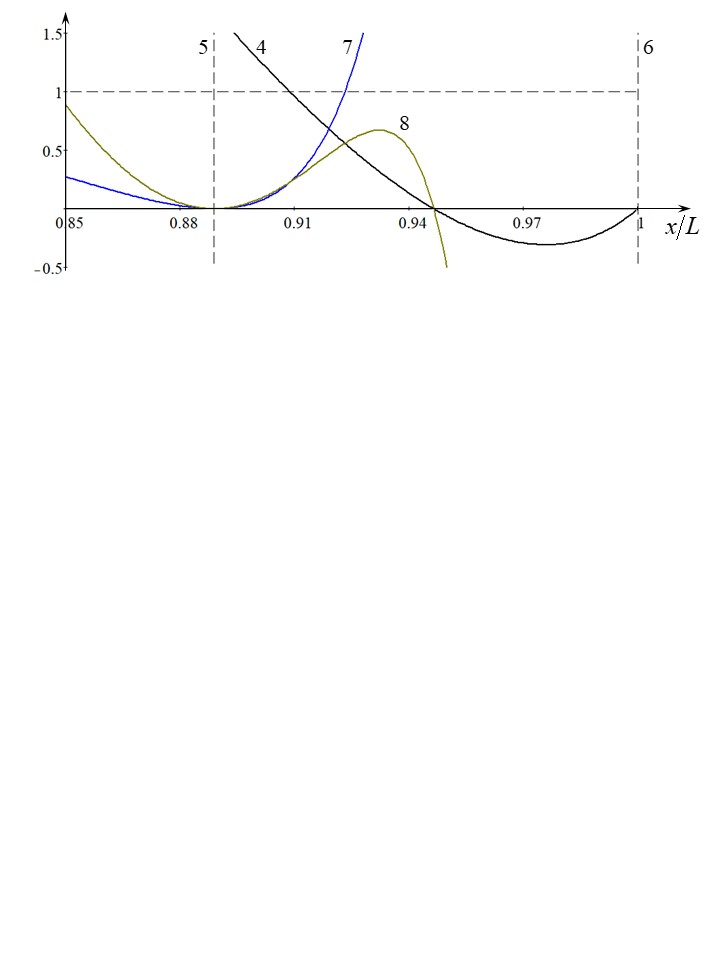}}} %
\vspace{-12.5cm}%
	\caption{\small A fragment of frame c) in Fig. \ref{f01}. Normalised coefficients of NLS equation (\ref{VarCoefNLS}): line 4 -- $50p(x)/p(0)$; line 7 -- $q(x)/q(0)$; line 8 -- $50p(x)q(x)/p(0)q(0)$; lines 5 and 6 are the same as in Fig. \ref{f01}.} %
	\label{f02}
\end{figure}

In the vicinity of distance $x = x_1 \approx 0.889L$ (see vertical dashed line 5) the dispersion and nonlinear coefficients experience dramatic changes. The nonlinear coefficient $q$ turns to zero at $x = x_1$, and then quickly increases on absolute value remaining negative for all $x \ne x_1$. The dispersion coefficient $p$ turns to zero little further, at $x_2 \approx 0.947L$, and then changes its sign from negative to positive; this is shown in detail in Fig. \ref{f02}, where the dispersion coefficient is multiplied by the factor 50 to make it more clearly visible. (We remind the readers again that these coefficients are presented in the normalised forms in Figs. \ref{f01} and \ref{f02}.) The same Fig. \ref{f02} also shows a normalised product $p(x)q(x)$ (see line 8) which determines the modulation stability/instability. The instability occurs when this product is positive (see, e.g. \cite{Karpman73, OstrPotap99, ZakhOstr09}). In our case the modulation instability providing the existence of NLS envelop solitons (\ref{EnvSol}) occurs when either $x < x_1$ or $x_1 < x < x_2$. In the vicinity of $x = x_1$ and $x = x_2$ the TNLS equation is not valid and should be replaced by the generalised NLS equation (\ref{GenNLS}). 

Now let us estimate the key parameters of the NLS envelope soliton (\ref{EnvSol}). A carrier frequency (where the growth rate of modulation instability has a maximum) is $\omega_{max} = c(9\gamma/4\beta)^{1/4} \approx 1.42\cdot 10^{-3}$ 1/s (the corresponding period is $T_{max} \approx 4.43\cdot 10^3$ s), and the maximal spatial growth rate (\ref{GammaMax}) for the initial amplitude $a_0 = 1$ m is $\Gamma_{max} \approx 1.94\cdot 10^{-5}$ 1/m. The carrier wave number $k(\omega_{max})$ where the growth rate of modulation instability has a maximum is $k_m \approx 1.23\cdot 10^{-3}$ 1/m (the corresponding wavelength $\lambda_m \approx 5.1$ km), whereas the carrier wave number $k(\omega_{m0})$ where the group velocity has a maximum is $k_c \approx 7.54\cdot 10^{-4}$ 1/m ($\lambda_c \approx 8.3$ km).

The nonlinear coefficient in the time-like NLS equation (\ref{VarCoefNLS}) at $x = 0$ is $q(k_m) = -2.7\cdot 10^{-5}$ 1/(m$^2$s), and the dispersion coefficient at the same point is $p(k_m) = -13.6$ (m$^2$/s).

\begin{figure}[h!]
	{\centerline{\includegraphics [scale=0.49]{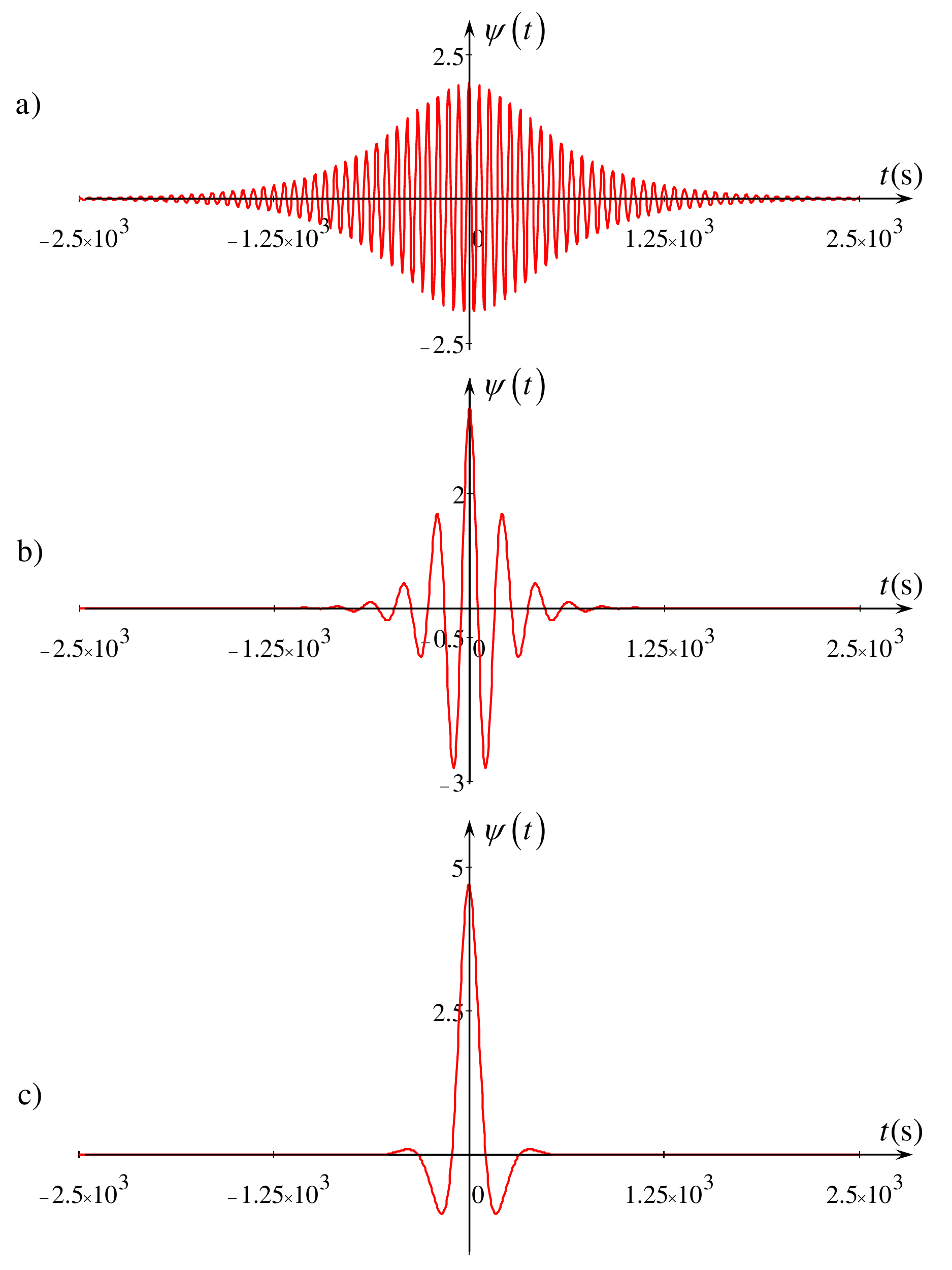}}} %
	\caption{\small An NLS envelope soliton (\ref{EnvSol}) at $x = 0$ (frame a), $x = 0.5L$ (frame b), and $x = 0.7L$ (frame c).} %
	\label{f03}
\end{figure}

If we assume that an NLS soliton (\ref{EnvSol}) has the initial amplitude $A_0 = 2$ m and velocity $V_0 = 0.5c_g(0) = 0.57$ m/s, then we obtain that its characteristic duration is: $\Delta T(0) \equiv 1/\Omega_0 = \sqrt{2p(k_m)/q(k_m)}/|A_0c_g(k_m)| \approx 440$ s $\approx 7.3$ min. The soliton chirp (\ref{EnvSolPar}) at $x = 0$ is $|\sigma(0)| = c_g(0)^3/2|p(0)|V(0) = 0.095$ 1/s and the corresponding carrier wave period is $T_c(0) \equiv 2\pi/|\sigma(0)| \approx 66$ s. This means that there are $N(0) = \Delta T(0)/T_c(0) \approx 7$ carrier wave periods on the half-duration of the envelope soliton (see frame a) in Fig. \ref{f03}).
\clearpage

\begin{figure}[t!]
	{\centerline{\includegraphics [scale=0.8]{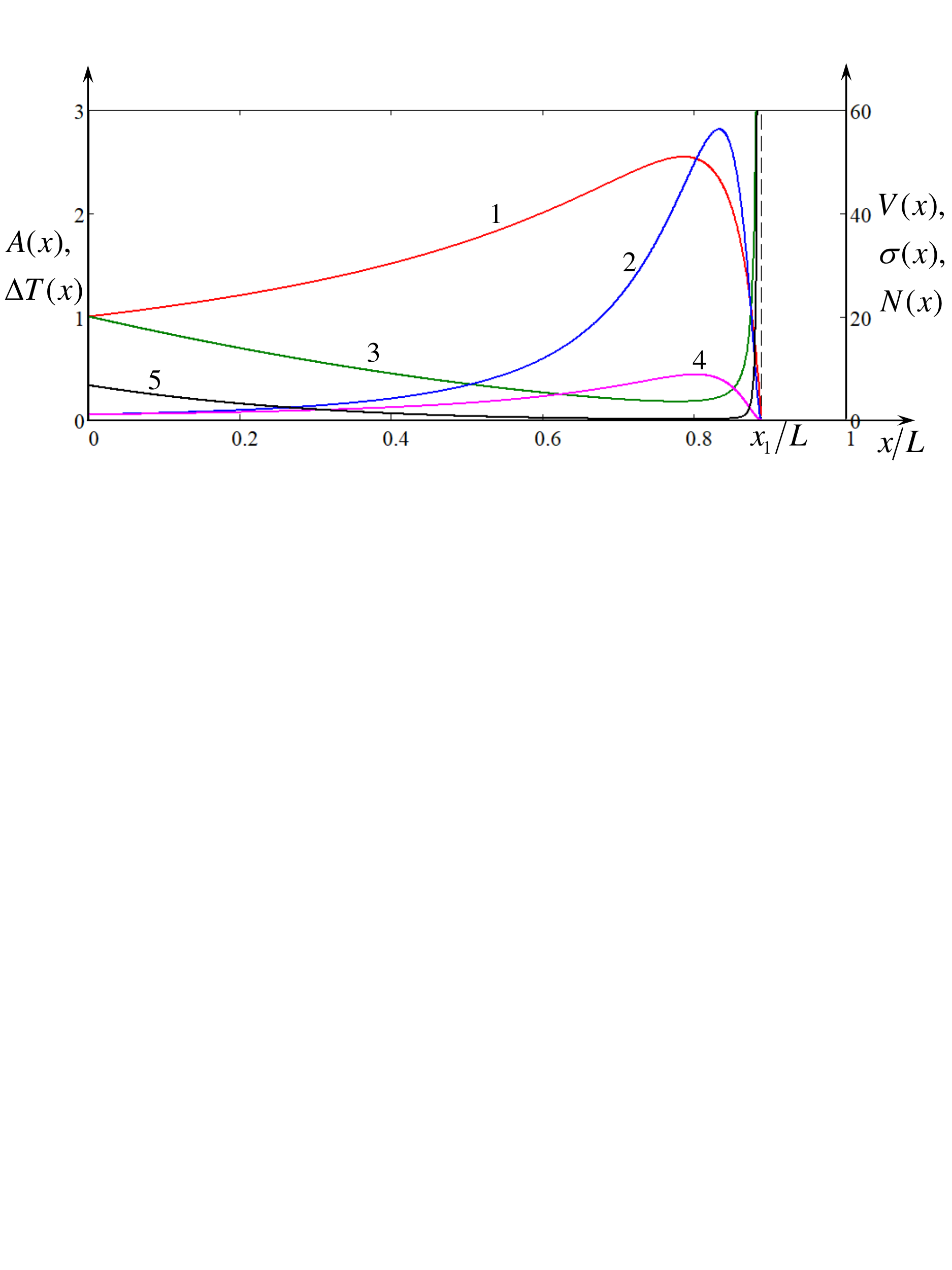}}} %
	\vspace{-13.0cm}%
	\caption{\small Variations with the distance of normalised soliton amplitude $A(x)/A(0)$ (line 1), velocity $V(x)/V(0)$ (line 2), duration $\Delta T(x)/\Delta T(0)$ (line 3), carrier wave period $T_c(x)/T_c(0)$ (line 4), and the number of carrier wave periods on the half-duration $N(x)$ (line 5) when a soliton moves towards the shore with the linearly decreasing depth as shown in Fig. \ref{f01}. } %
	\label{f04}
\end{figure}

In the process of propagation towards the shoaling, a soliton experiences amplitude enhancement as per Eq. (\ref{Eq1}) with simultaneous shrinking of its duration; its velocity $V$ also increases with $x$ in accordance with Eq. (\ref{Eq3}). Variation of soliton amplitude, velocity, duration, as well as carrier wave period $T_c(x) = 2\pi/\sigma(x)$ are shown in Fig. \ref{f04}. As one can see from this figure, both the amplitude and velocity increase first and attain the maximal values, but at different distances; after that they quickly decrease when the soliton approaches the distance where the lower and upper layers have equal thicknesses, and the nonlinear coefficient turns to zero (see lines 1 and 2 in  Fig. \ref{f04}). In particular, the maximal soliton amplitude becomes 2.55 times greater than the initial one, whereas the maximal velocity becomes 57 times greater than the initial one. Note that a very similar effect of soliton amplitude is that it decrease (while the duration increases) upon approaching the critical depth where $kh_c = 1.363$ was revealed in Ref. \cite{Benilov05} when the soliton propagates in a non-rotating fluid with a smoothly varying bottom. In both cases (in this paper and in Ref. \cite{Benilov05}) the reason for the decrease in soliton amplitude is in the vanishing of the nonlinear coefficients in the NLS equations, although they vanish at different values of $kh$.

When $x$ becomes greater than $x_1$, the soliton amplitude and velocity formally increase again and go to infinity when $x \to x_2$ (this is not shown in the plot). However, the adiabatic theory breaks much earlier, when $x \to x_1$, and the process of wave transformation in the vicinity of this point should be reconsidered more thoroughly, apparently, on the basis of primitive equations. 

The soliton duration varies inverse proportional to the amplitude; therefore it decreases first, but then dramatically increases when $x \to x_1$ (see line 3 in Fig. \ref{f04}). In contrast, the period of the carrier wave increases first, attains a maximal value and then quickly drops to zero, when $x \to x_1$ see line 4 in Fig. (\ref{f04}). As a result of this, the number of carrier wave periods within the soliton half-duration $N(x)$ decreases first up to zero and then formally increase again when $x \to x_1$ (see line 5 in Fig. \ref{f04}).

Transformation of the soliton shape in the process of its propagation towards the cost is illustrated by Fig. \ref{f03}. The soliton shape at the half-way to the coasts when $x = 0.5L$ is still close to the NLS envelope soliton (see frame b). However at $x = 0.7L$ where the soliton amplitude becomes close to the maximal value, it looks rather like a solitary wave, but represents a nonstationary formation oscillating in time (see frame c). Such a formation can hardly be described by the TNLS equation which presumes that $N(x) >> 1$.

Thus, in the process of soliton propagation towards the shoaling, it becomes narrower and transforms from the wave train shown in frame a) of Fig. \ref{f03} to the pulse-type nonstationary solitary wave (a breather) as shown in frame c). To certain extent, this behaviour is opposite to the transformation of the KdV solitary wave into an envelope soliton described, for example, in Refs. \cite{Grimshaw08, Grimshaw12, Grimshaw13, Grimshaw16}.

\section{Discussion and conclusion}
\label{Sect08}

In this paper it has been demonstrated that in the rotating fluid the most probable frequency of the carrier wave which constitutes the NLS solitary wave is the frequency where the growth rate of modulation instability is maximal. This agrees with the conjecture of Ref. \cite{wj15} and the numerical results of that paper. This frequency differs from the frequency where the group velocity has a maximum as was originally hypothesised in Refs. \cite{Grimshaw08,Grimshaw12,Grimshaw13}. Envelope solitary wave of this frequency can be described by the conventional TNLS equation (\ref{VarCoefNLS}), rather than the generalised NLS equtaion (\ref{GenNLS}). Soliton solutions to both these equations have been presented and the limiting cases when some coefficients vanish have been discussed.

If an internal envelope soliton has been formed in a homogeneous two-layer rotating ocean and then enters a coastal zone, where the bottom linearly increases with a small gradient, then it experiences an adiabatic transformation. This leads to an increase of soliton amplitude, velocity, and period of a carrier wave, whereas its duration decreases. Therefore, it becomes taller and narrower. At some distance it looks like a breather, i.e., a narrow nonstationary solitary wave. Apparently, the TNLS equation is not quite appropriate for the description of its further evolution; more advanced theory and/or numerical simulation are required to this end. This can be a theme for a further study. \\

{\bf Acknowledgements.} The author acknowledges the funding of this study from the State task program in the sphere of scientific activity of the Ministry of Education and Science of the Russian Federation (Project No. 5.1246.2017/4.6) and grant of the President of the Russian Federation for state support of leading scientific schools of the Russian Federation (NSH-2685.2018.5).

\bibliographystyle{plain}

\end{document}